%% file: main.tex
\documentclass[conference, a4paper]{IEEEtran}

\renewcommand\IEEEkeywordsname{Index Terms}
\usepackage{graphicx}
\usepackage{subcaption}
\usepackage{tikz,xparse}

\usetikzlibrary{patterns}

\usetikzlibrary{dsp,chains}
\usetikzlibrary{matrix}
\usepackage{mathptmx}
\usepackage{verbatim}
\usepackage{calc}
\usepackage{ifthen}
\usepackage{xifthen}
\usepackage{cancel}
\usepackage{bm}
\usepackage{verbatim}
\usepackage{multirow}
\usepackage{cite}
\usepackage[hyphens]{url}
\usepackage[nolist]{acronym} 
\usepackage{pgfplots}
\usetikzlibrary{arrows,shapes,graphs,graphs.standard,quotes,arrows.meta,decorations.markings,positioning}

\pgfplotsset{compat=newest}
\usepackage[bookmarks=false]{hyperref}
\usepackage{units}
\usepackage{amsmath, amsbsy, amssymb, latexsym}
\DeclareMathAlphabet{\mathcal}{OMS}{cmsy}{m}{n}
\usepackage{mathtools}
\hypersetup{bookmarksdepth=-2}
\usepackage{comment}
\usepackage[utf8]{inputenc}
\usepackage{xcolor}
\usepackage{enumitem}
\usepackage[linesnumbered,vlined, ruled]{algorithm2e}
\usepackage{algpseudocode}
\usepackage{etoolbox} 
\usepackage[normalem]{ulem}

\usepackage{marginnote}
\tikzset{>=latex}
\SetAlCapNameFnt{\footnotesize}
\SetAlCapFnt{\footnotesize}


\input{corporateColours.tex}
\input{nodesAndStyles.tex}

\newcommand{\qed}{\hfill\blacksquare}

\IEEEoverridecommandlockouts

\input{macros}		

\begin{document}

\begin{NoHyper}
\title{Rate-Compatible Polar Codes for Automorphism Ensemble Decoding}

\author{\IEEEauthorblockN{Marvin Geiselhart, Jannis Clausius and Stephan ten Brink}
	\IEEEauthorblockA{
		Institute of Telecommunications, Pfaffenwaldring 47, University of  Stuttgart, 70569 Stuttgart, Germany 
		\\\{geiselhart,clausius,tenbrink\}@inue.uni-stuttgart.de\\
	}
		\thanks{This work is supported by the German Federal Ministry of Education and Research (BMBF) within the project Open6GHub (grant no. 16KISK019).}
  }

\makeatletter
\patchcmd{\@maketitle}  
{\addvspace{0.5\baselineskip}\egroup}
{\addvspace{-1.0\baselineskip}\egroup}
{}
{}
\makeatother

\maketitle

\begin{acronym}
\acro{ML}{maximum likelihood}
\acro{BP}{belief propagation}
\acro{BPL}{belief propagation list}
\acro{LDPC}{low-density parity-check}
\acro{BER}{bit error rate}
\acro{SNR}{signal-to-noise-ratio}
\acro{BPSK}{binary phase shift keying}
\acro{AWGN}{additive white Gaussian noise}
\acro{LLR}{Log-likelihood ratio}
\acro{MAP}{maximum a posteriori}
\acro{FER}{frame error rate}
\acro{BLER}{block error rate}
\acro{SCL}{successive cancellation list}
\acro{SC}{successive cancellation}
\acro{BI-DMC}{Binary Input Discrete Memoryless Channel}
\acro{CRC}{cyclic redundancy check}
\acro{CA-SCL}{CRC-aided successive cancellation list}
\acro{BEC}{Binary Erasure Channel}
\acro{BSC}{Binary Symmetric Channel}
\acro{BCH}{Bose-Chaudhuri-Hocquenghem}
\acro{RM}{Reed--Muller}
\acro{RS}{Reed-Solomon}
\acro{SISO}{soft-in/soft-out}
\acro{3GPP}{3rd Generation Partnership Project }
\acro{eMBB}{enhanced Mobile Broadband}
\acro{CN}{check node}
\acro{VN}{variable node}
\acro{GenAlg}{Genetic Algorithm}
\acro{CSI}{Channel State Information}
\acro{OSD}{ordered statistic decoding}
\acro{MWPC-BP}{minimum-weight parity-check BP}
\acro{FFG}{Forney-style factor graph}
\acro{MBBP}{multiple-bases belief propagation}
\acro{URLLC}{ultra-reliable low-latency communications}
\acro{DMC}{discrete memoryless channel}
\acro{SGD}{stochastic gradient descent}
\acro{QC}{quasi-cyclic}
\acro{5G}{fifth generation mobile telecommunication}
\acro{SCAN}{soft cancellation}
\acro{LSB}{least significant bit}
\acro{AED}{automorphism ensemble decoding}
\acro{AE-SC}{automorphism ensemble successive cancellation}
\acro{PPV}{Polyanskyi-Poor-Verd\'{u}}
\end{acronym}

\begin{abstract}
Recently, \ac{AED} has drawn research interest as a more computationally efficient alternative to \ac{SCL} decoding of polar codes. Although \ac{AED} has demonstrated superior performance for specific code parameters, a flexible code design that can accommodate varying code rates does not yet exist. This work proposes a theoretical framework for constructing rate-compatible polar codes with a prescribed automorphism group, which is a key requirement for \ac{AED}. We first prove that a one-bit granular sequence with useful automorphisms cannot exist. However, by allowing larger steps in the code dimension, flexible code sequences can be constructed. An explicit synthetic channel ranking based on the $\beta$-expansion is then proposed to ensure that all constructed codes possess the desired symmetries. Simulation results, covering a broad range of code dimensions and blocklengths, show a performance comparable to that of 5G polar codes under \ac{CRC}-aided \ac{SCL} decoding, however, with lower complexity.

\end{abstract}
\acresetall



\section{Introduction}
Polar codes, introduced by Arıkan \cite{ArikanMain}, are the first class of channel codes that provably asymptotically achieve the channel capacity of a \ac{DMC} under low-complexity \ac{SC} decoding for long blocklengths.
To improve their performance in the short blocklength regime, an outer \ac{CRC} code and the \ac{SCL} decoder have been introduced \cite{talvardyList}. For this reason, \ac{SCL} has become the de-facto standard decoder for \ac{CRC}-aided polar codes, which have been adopted in the 5G standard \cite{polar5G2018}. 

Unfortunately, \ac{SCL} decoding requires significant path management overhead, which increases decoding latency and makes implementations with large list sizes difficult.
An alternative to \ac{SCL} decoding is \ac{AED} \cite{rm_automorphism_ensemble_decoding}, where all sub-decoders are independent, allowing much more efficient hardware implementations \cite{Kestel2023}.
\ac{AED} uses the automorphism group of a code to generate a set of $M$ permuted versions of the received sequence, which are processed in parallel by low-complexity \ac{SC} decoders. Thus, for \ac{AED} to be applicable, the code must have a reasonable number of useful automorphisms, which requires special code design \cite{polar_aed}. 
The first code design for \ac{AED} was proposed in \cite{pilletPolarCodesForAED}, but it is guided by human inspection and is therefore not fully automatic. 
In \cite{pillet2022AEDdesign}, the first automated code design is presented that ensures a prescribed automorphism group. This is achieved by iteratively modifying a polar code generated by a standard design algorithm until it has the desired symmetries.
A more direct approach is proposed in \cite{ShabunovPredefinedAutomorphisms}, where information bits are added to an arbitrary design until the symmetry constraints are satisfied.
All of the above design algorithms produce a code with only a single code rate, but many practical applications require codes that can accommodate different rates. 
In \cite{pillet2022distribution} it is shown that highly symmetric codes exist for almost all code dimensions. However, these codes are not subcodes of each other and therefore do not form a rate-compatible sequence.
In this paper, we propose a low-complexity, explicit design algorithm for rate-flexible polar codes.
The main contributions of this paper are as follows:
\begin{itemize}
    \item We formulate necessary and sufficient conditions on the information sets of polar code sequences to contain at least a predefined automorphisms group.
    \item We extend the partial order to include exactly the codes with at least this symmetry.
    \item Based on the $\beta$-expansion, we propose a bit reliability sequence that inherently captures the desired symmetries and, thus, directly yields rate-flexible polar code designs suitable for \ac{AED}.
    \item For the first time, close to state-of-the-art performance of \ac{AED} is demonstrated over a broad range of code rates and blocklengths with extensive Monte-Carlo simulation.
\end{itemize}

\section{Preliminaries}\label{sec:preliminaries}

\subsection{Notation}
Vectors and matrices are indicated by lowercase and uppercase boldface letters, respectively.
$\mathbb{Z}_N$ denotes the set of integers between 0 and $N-1$.
For an integer $i\in \mathbb{Z}_{2^n}$, $\hat{\iv} \in \mathbb{F}_2^n$ denotes its $n$-bit \ac{LSB}-first binary expansion. Binary numbers are written \ac{LSB}-first and indicated by the subscript 2, e.g., $11=11010_2$. 
$S_n$ denotes the set of all permutations of $n$ elements.

\subsection{Polar Codes}
The structure of a polar code is defined by the $N\times N$ Hadamard matrix $\mathbf G_N = \left[\begin{smallmatrix} 1 & 0 \\ 1 & 1 \end{smallmatrix}\right]^{\otimes n}$ with $N=2^n$, where $(\cdot)^{\otimes n}$ denotes the $n$-th Kronecker power.
The Hadamard matrix can be interpreted as a transformation of $N$ \acp{DMC} into $N$ polarized synthetic channels. 
In this context, \textit{polarized} means that the majority of synthetic channels are either highly reliable or highly unreliable, with only a few being intermediate
The task of polar code design is to select a subset of reliable channel indices for information transmission, the so called \emph{information set} $\mathcal{I}$. 
The remaining, unreliable channel indices form the so called \emph{frozen set} $\mathcal{F}$, where frozen bits equal to ``0'' are transmitted.
The corresponding generator matrix $\mathbf{G}$ is composed of the rows of $\mathbf G_N $ defined by $\mathcal{I}$. 

\subsection{Successive Cancellation Decoding}
The \ac{SC} decoding algorithm works by depth-first traversal of the factor graph. The bits associated with the synthetic channels are successively estimated based on the channel observations and the hard decisions of all previously estimated bits. At frozen positions, the hard decision is always set to 0. In the following, we denote the \ac{SC} decoding function by $\tilde{\xv}=\operatorname{SC}(\yv)$, where $\yv$ is the received sequence and $\tilde{\xv}$ is the codeword estimate.
\ac{SCL} modifies the depth-first traversal of the factor graph by branching at each information bit into two parallel decoding paths, assuming the bit to be 0 and 1 respectively.
In addition, a path metric is calculated based on the reliability of the bits.
To counteract the exponential growth of the number of paths, only the $L$ paths with the smallest path metrics are kept at each branching step \cite{talvardyList}.

\subsection{Partial Order}

The synthetic channels of a polar code display a partial order in a sense that some are always more reliable than others, independent of the transmission channel \cite{bardet_polar_automorphism}, \cite{partialorder}. 
We write $i \preccurlyeq j$, if the synthetic channel $j$ is always at least as reliable as $i$. The partial order is defined by two rules\footnote{In \cite{bardet_polar_automorphism} the partial order is defined using the monomial formalism, which can be easily related to the notation used here. A formal proof of their equivalence is given in \cite{bioglio2023groupproperties}.}:
\begin{enumerate}
    \item \textit{Left swap}: If $(\hat{i}_l,\hat{i}_{l+1})=(1,0)$, and  $(\hat{j}_l,\hat{j}_{l+1})=(0,1)$ for some $l\in \ZZ_{n-1}$, while all other $\hat{j}_{l'}=\hat{i}_{l'}$, $l'\not\in\{l,l+1\}$, then $i \preccurlyeq j$. 
    That is, if $\hat{\jv}$ equals $\hat{\iv}$ with a single ``1" bit moved one position to the right (more significant), then $i \preccurlyeq j$. 
    \item \textit{Binary domination}: If $\hat{i}_l\le\hat{j}_l$ $\forall l\in \ZZ_n$, then $i \preccurlyeq j$. 
    In other words, $\hat{\jv}$ is $\hat{\iv}$, but may have additional ``1" bits.
\end{enumerate}
All remaining relations can be derived from transitivity.

A polar code with information set $\mathcal{I}$ complies with (or follows) the partial order if
$i \in \mathcal{I} $ implies that all $j \succcurlyeq i$ also satisfy $j \in \mathcal{I}$,
i.e., for any information bit, all more reliable channels are also designated as information bits.

A low-complexity polar code design algorithm that always follows the partial order is the $\beta$-expansion \cite{BetaIngmard}. It assigns each synthetic channel a \textit{polarization weight} computed as
\begin{equation} 
    w_\mathrm{p}(i) = \sum_{l=0}^{n-1} \hat{i}_l \beta^l \label{eq:betaexp}
\end{equation}
where $\beta$ is a real number satisfying $1 < \beta \le 2$. The synthetic channels are ranked by their polarization weight and the $K$ with largest weight are selected as the information set. 

\subsection{Automorphisms of Polar Codes}
\textbf{Definition:} 
The (permutation) automorphism group $\operatorname{Aut}(\mathcal{C})$ of a code $\mathcal{C}$ with blocklength $N$ is the set of all permutations of the codeword symbols that map each codeword onto another (not necessarily different) codeword. In other words,
\begin{equation*}
    \operatorname{Aut}(\mathcal{C}) = \left\{ \pi \in S_N | \pi(\mathbf{c}) \in \mathcal{C}\; \forall \mathbf{c} \in \mathcal{C} \right\}.
\end{equation*}

In the context of polar codes, a special form of permutations is particularly interesting, namely \textit{affine permutations}.

\textbf{Definition:} 
An affine permutation is a permutation of the codeword symbols that can be written as
\begin{equation}
    \pi: i\mapsto i': \quad \hat{\iv}' = \Am\hat{\iv}+\bv \label{eq:affine_perm}
\end{equation}
where $\mathbf{A}\in \mathbb{F}_2^{n\times n}$ is a non-singular matrix and $\mathbf{b}\in \mathbb{F}_2^n$. 
Similarly, the affine automorphism group of a code is the subgroup of automorphisms that are affine permutations.

For polar codes, the affine automorphism group can be characterized by the notion of stabilizers:

\textbf{Definition:} 
The stabilizer $\operatorname{Stab}(\mathcal{S})$ of a set $\mathcal{S}\subseteq \mathbb{Z}_{2^n}$ is the set of all permutations $\sigma \in S_n$ with the property
\begin{equation*}
    f_\sigma(i) \in \mathcal{S} \quad \forall i \in \mathcal{S},
\end{equation*}
where $f_\sigma(i)$ permutes the binary digits of $i$, i.e.,
\begin{equation*}
    f_\sigma: i \mapsto i': \quad \hat{\iv}' = \sigma(\hat{\iv}). 
\end{equation*}

If $\mathcal{I}$ is the information set of a polar code following the partial order, then $\operatorname{Stab}(\mathcal{I})$ follows a block structure, i.e., bits are only permuted within contiguous blocks \cite{polar_aed}. This block structure is defined by the vector of block sizes $\mathbf{s} = [s_0, ..., s_{m-1}] $, with $m$ being the number of blocks. We define this set $P_\mathbf{s}\subseteq S_n$ as those permutations whose permutation matrices have no non-zero entries above the block profile $\mathbf{s}$.

\textbf{Theorem 1:} Let $\mathcal{C}$ be a polar code with information set $\mathcal{I}$ following the partial order and $\mathbf{s}$ being the block profile of $\operatorname{Stab}(\mathcal{I})$.
Then, its affine automorphism group is the set of affine permutations $\operatorname{BLTA}(\mathbf{s})$ according to equation (\ref{eq:affine_perm}) where $\mathbf{A}$ does not have any non-zero elements above the block diagonal given by $\mathbf{s}$ \cite{polar_aed, liBLTA2021}.

In the following, such a code is called $\sv$-\textit{symmetric}.

\subsection{Automorphism Ensemble Decoding}
\ac{AED} has been proposed in \cite{rm_automorphism_ensemble_decoding} as a highly parallelizable decoding algorithm for \ac{RM} codes, which has been applied to polar codes in \cite{polar_aed}. It uses an ensemble of $M$ identical low-complexity \ac{SC} decoders
that all run in parallel on permuted versions of the noisy received sequence $\yv$. Then, the decoding results are depermuted to
\begin{equation*}
    \tilde{\xv}_j = \pi_j^{-1}(\operatorname{SC}(\pi_j(\yv)))
\end{equation*}
from which the most likely candidate is picked as the final codeword estimate
\begin{equation*}
    \tilde{\xv} = \underset{\tilde{\xv}_j, j = 0,\dots,M-1}{\operatorname{arg\,max}} P( \yv |\tilde{\xv}_j).
\end{equation*}
The permutations $\pi_j$ stem from the automorphism group of the code. Due to decoder equivariance \cite{rm_automorphism_ensemble_decoding}, at most one from each equivalence class is used in \ac{AED} \cite{pillet2022AEDdesign}.

\section{Polar Codes with Symmetry Constraints}
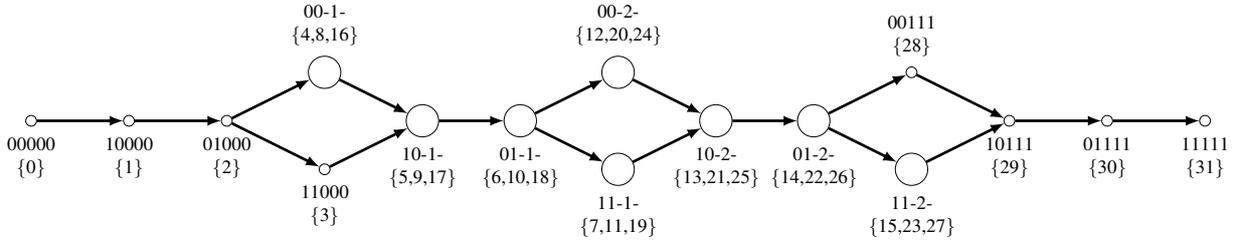
\begin{figure*}
    \centering
    \resizebox{.9\linewidth}{!}{\input{tikz/symmetric_partial_order}}
    \caption{\footnotesize Hasse diagram of the $[1,1,3]$-symmetric partial order for $N=32$ polar codes, with binary and base-10 indices. Each of the large nodes corresponds to a group of three synthetic channel indices, where ``-$x$-'' denotes that any combination of $x$ ``1s'' appears in these 3 bits of the index.}
    \label{fig:symmetric_partial_order}
\end{figure*}

\subsection{Symmetric Code Augmentation}
In the following, we derive relationships between codes that have at least a prescribed set of automorphisms $\operatorname{BLTA}(\mathbf{s})$. 
Our approach is similar to that of \cite{Moscow_Huawei_Polar_paper, ShabunovPredefinedAutomorphisms}, where groups of indices that are either entirely frozen or information bits are identified. 
However, as we are interested in the full-affine automorphism group of polar codes, we additionally consider the partial order of synthetic channels. To analyze which information bits can be added to a code while preserving its symmetry, we utilize again the concept of stabilizers.

\textbf{Proposition 1:} Let $\mathbf{s}$ be some block profile, and a $\mathcal{I}_1$ an information set compliant with the partial order and $\operatorname{Stab}(\mathcal{I}_1) \supseteq P_\mathbf{s}$. Moreover, let $\mathcal{I}_2 = \mathcal{I}_1 \cup \mathcal{D}$ also follow the partial order with $\mathcal{D}\cap \mathcal{I}_1=\varnothing$. Then $\operatorname{Stab}(\mathcal{I}_2) \supseteq P_\mathbf{s}$ if and only if $\operatorname{Stab}(\mathcal{D}) \supseteq P_\mathbf{s}$.

\textit{Proof.} ``$\Rightarrow$'': Suppose $\operatorname{Stab}(\mathcal{D}) \not\supseteq P_\mathbf{s}$, i.e., $\exists i\in \mathcal{D}, \sigma \in P_\mathbf{s}$ with $f_\sigma(i) \not\in \mathcal{D}$. Then $i \in \mathcal{I}_2$ but $f_\sigma(i) \not\in \mathcal{I}_2$, which contradicts $\operatorname{Stab}(\mathcal{I}_2) \supseteq P_\mathbf{s}$. Note that $f_\sigma(i) \in \mathcal{I}_1$ is not possible, as then also $i \in \mathcal{I}_1$ to ensure $\operatorname{Stab}(\mathcal{I}_1) \supseteq P_\mathbf{s}$.

``$\Leftarrow$'': This directly follows from $\operatorname{Stab}(\mathcal{I}_2) = \operatorname{Stab}(\mathcal{I}_1 \cup \mathcal{D}) = \operatorname{Stab}(\mathcal{I}_1) \cap \operatorname{Stab}(\mathcal{D})$ which is at least $P_\mathbf{s}$. $\qed$

This means that if two $\sv$-symmetric polar codes are subcodes of each other, their difference index set must follow the partial order and be itself stable under $P_\mathbf{s}$.
Clearly, this restricts the granularity of of the code dimension $K$ of a sequence of symmetric rate-compatible polar codes. For the finest granularity, let $\mathcal{D}_i$ denote the set of channel indices that contains only $i$ and all its permutations under $P_\mathbf{s}$, i.e.,
$\mathcal{D}_i = \left\{f_\sigma(i) | \sigma \in P_\mathbf{s} \right\}$.
The size of $\mathcal{D}_i$ is computed as
\begin{equation*}
    |\mathcal{D}_i| = \prod_{j=0}^{m-1} \binom{s_j}{w_{\mathrm{h},\mathbf{s},j}(\hat{\iv})}, 
\end{equation*}
where $w_{\mathrm{h},\mathbf{s},j}(\hat{\iv})$ denotes Hamming weight of $\hat{\iv}$ in the $j$-th block of $\mathbf{s}$.
For $n=4$, Table \ref{tab:example} shows the groups stabilized by $P_\sv$ with $\mathbf{s}=[2,2]$. For example, for $i=5=1010_\mathrm{2}$, we have $\mathcal{D}_i~=~\{1010_\mathrm{2},\,0110_\mathrm{2},\,1001_\mathrm{2},0101_\mathrm{2}\}=\{5,6,9,10\}$ of size 4.

\subsection{Symmetric Partial Order}
The partial order introduced in \cite{bardet_polar_automorphism}, \cite{partialorder} can be restricted to ensure that all constructed polar codes attain at least $\operatorname{BLTA}(\mathbf{s})$ as their automorphism group, for a given $\sv$. As discussed above, a synthetic channel $i$ can only be added or removed together with all elements of its group $\mathcal{D}_i$ which is stable under $P_\mathbf{s}$.
Hence, we define a partial order ``$\preccurlyeq_\sv$'' of the groups as
\begin{equation*}
    \mathcal{D}_i \preccurlyeq_\sv \mathcal{D}_j \quad \Leftrightarrow  
    \quad \exists i' \in \mathcal{D}_i, j' \in \mathcal{D}_j \text{ with } i' \preccurlyeq j'.
\end{equation*}
The Hasse diagram of ``$\preccurlyeq_\sv$'' can be derived from the one of ``$\preccurlyeq$'' by merging the nodes of synthetic channels within a group stable under $P_\mathbf{s}$ into a single node.
Observe that this merging can always be done, as all elements of each group are comparable under the partial order and directly neighbor each other (left swap rule). In other words,
\begin{equation*}
    \mathcal{D}_i \preccurlyeq_\sv \mathcal{D}_j \quad \Rightarrow \quad \forall j' \in \mathcal{D}_j \quad \exists i' \in \mathcal{D}_i \text{ with } i' \preccurlyeq j'.
\end{equation*}
Fig.~\ref{fig:symmetric_partial_order} shows the Hasse diagram for $n=5$, and a minimum symmetry of $\mathbf{s}=[1,1,3]$.

\textbf{Theorem 2:} Let $\mathbf{s}$ be some block profile. A polar code $\mathcal{C}$ satisfies $\operatorname{Aut}(\mathcal{C}) \supseteq \operatorname{BLTA}(\mathbf{s})$ if and only if its information set complies with ``$\preccurlyeq_\sv$''.

\textit{Proof.} ``$\Rightarrow$'': We have $P_\mathbf{s}$ stabilizing $\mathcal{I}$ and, as per \cite{bioglio2023groupproperties}, Thm. 3, we know that $\mathcal{C}$ complies with ``$\preccurlyeq$'' if $\operatorname{Aut}(\mathcal{C}) \supseteq \operatorname{BLTA}(\mathbf{s})$, and, hence, it also follows ``$\preccurlyeq_\sv$''.

``$\Leftarrow$'': By the definition above, every code complying with ``$\preccurlyeq_\sv$'', also follows the partial order ``$\preccurlyeq$'' and is stabilized by $P_\mathbf{s}$. By Thm. 1, we therefore have $\operatorname{Aut}(\mathcal{C}) \supseteq \operatorname{BLTA}(\mathbf{s})$. $\qed$

\section{Rate-Compatible Sequences of Symmetric Polar Codes}
\begin{table}
    \centering
    \caption{\footnotesize Example for $n=4$, $\sv=\left[2,2\right]$. Grouped entries indicate the different stabilizing subsets $\mathcal{D}_i$.}
    
    \begin{tabular}{c c|c c|c c|c c}
        $i$ & $\hat{\iv}$ & $i$ & $\hat{\iv}$ & $i$ & $\hat{\iv}$ & $i$ & $\hat{\iv}$ \\
        \hline
        0 & \textcolor{black}{0000\textsubscript{2}}& 4 & \textcolor{mittelblau}{00\textbf{10}\textsubscript{2}} & 8 & \textcolor{mittelblau}{00\textbf{01}\textsubscript{2}} & 12 & 0011\\
        1 & \textcolor{rot}{\textbf{10}00\textsubscript{2}} & 5 & \textcolor{apfelgruen}{\textbf{1010}\textsubscript{2}}  & 9 & \textcolor{apfelgruen}{\textbf{1001}\textsubscript{2}} &13 & \textcolor{orange}{\textbf{10}11\textsubscript{2}}  \\
        2 & \textcolor{rot}{\textbf{01}00\textsubscript{2}} &6 & \textcolor{apfelgruen}{\textbf{0110}\textsubscript{2}}  & 10 & \textcolor{apfelgruen}{\textbf{0101}\textsubscript{2}}& 14 & \textcolor{orange}{\textbf{01}11\textsubscript{2}} \\
        3 & 1100\textsubscript{2} & 7 & \textcolor{violet}{11\textbf{10}\textsubscript{2}} & 11 & \textcolor{violet}{11\textbf{01}\textsubscript{2}} & 15 & 1111\textsubscript{2}
    \vspace{-1.65cm}  
    \end{tabular}
    \include{tikz/table_overlay}
    \label{tab:example}
    \vspace{-0.4cm}
\end{table}

Using the symmetric partial order, polar codes with a prescribed symmetry can be constructed. 
In applications that require multiple code rates, it is helpful to define a total order of the synthetic channels, rather than only a partial order. 
Inspired by the $\beta$-expansion, we propose an $\mathbf{s}$-symmetric polarization weight $w_{\mathrm{p},\mathbf{s}}$ with the following requirements:
\begin{enumerate}
    \item Equality: Channel indices in the same group should have the same polarization weight, i.e.,
    \begin{equation}
        w_{\mathrm{p},\mathbf{s}}(i) = w_{\mathrm{p},\mathbf{s}}(f_\sigma(i))\quad \forall \sigma \in P_\mathbf{s} \label{eq:similarity}
    \end{equation}
    \item Compatibility: The value of this polarization weight should lie between the minimum and maximum polarization weight of the original $\beta$-expansion, i.e.,
    \begin{equation}
        \min\limits_{i'\in \mathcal{D}_i} w_\mathrm{p}(i') \le w_{\mathrm{p},\mathbf{s}}(i) \le \max\limits_{i'\in \mathcal{D}_i} w_\mathrm{p}(i')  \label{eq:compatibility}
    \end{equation}
\end{enumerate}
For a similarly elegant formulation as eq. (\ref{eq:betaexp}), we only relax the condition that each bit position $l$ is associated to a power of $\beta$, but rather a general value $\beta_{\sv,l}$, leading to
\begin{equation*}
    w_{\mathrm{p},\sv}(i) = \sum_{l=0}^{n-1} \hat{i}_l \beta_{\sv,l}.
\end{equation*}
To satisfy (\ref{eq:similarity}), $\beta_{\mathbf{s},l}$ must be the same value within each block in $\mathbf{s}$. For condition (\ref{eq:compatibility}) to hold, also groups with only a single index with all-ones in a block (e.g., $i=N-1$) must satisfy $w_{\mathrm{p},\mathbf{s}}(i) = w_{\mathrm{p}}(i)$. Hence, the only valid definition for $\beta_{\mathbf{s},l}$ is 
\begin{equation*}
    \beta_{\mathbf{s},l} = \frac{1}{|\mathcal{L}_{\mathbf{s},l}|} \sum_{l' \in \mathcal{L}_{\mathbf{s},l}} \beta^{l'},
\end{equation*}
where $\mathcal{L}_{\mathbf{s},l}$ is the set of indices corresponding to the same block in $\mathbf{s}$ as the index $l$.

\textbf{Lemma 1}: For all $\beta \ge 1$ and $\sv$, we have $\beta_{\sv,l} \le \beta_{\sv,l+1}$.

\textit{Proof.} We can upper bound $\beta_{\sv,l}$ by
\begin{equation*}
    \beta_{\sv,l} = \frac{1}{|\mathcal{L}_{\mathbf{s},l}|} \sum_{l' \in \mathcal{L}_{\mathbf{s},l}} \beta^{l'} \le \frac{1}{|\mathcal{L}_{\mathbf{s},l}|} |\mathcal{L}_{\mathbf{s},l}| \cdot \beta^{\max\{\mathcal{L}_{\mathbf{s},l}\}} = \beta^{\max\{\mathcal{L}_{\mathbf{s},l}\}} 
\end{equation*}
and similarly lower bound $\beta_{\sv,l+1}$ by
\begin{equation*}
    \beta_{\sv,l+1} = \frac{1}{|\mathcal{L}_{\mathbf{s},l+1}|} \sum_{l' \in \mathcal{L}_{\mathbf{s},l+1}} \beta^{l'} \ge \beta^{\min\{\mathcal{L}_{\mathbf{s},l+1}\}}.
\end{equation*}
As $l'\le l''$ holds for all $l' \in \mathcal{L}_{\mathbf{s},l}$ and $l'' \in \mathcal{L}_{\mathbf{s},l+1}$, we have $ \beta_{\sv,l} \le \beta^{\max\{\mathcal{L}_{\mathbf{s},l}\}} \le \beta^{\min\{\mathcal{L}_{\mathbf{s},l+1}\}} \le \beta_{\sv,l+1}$.$\qed$

\textbf{Lemma 2}: Let $\mathbf{s}$ be some block profile and $i, j\in \ZZ_N$ distinct and following $i\preccurlyeq j$. Then also $w_{\mathrm{p},\mathbf{s}}(i) \le  w_{\mathrm{p},\mathbf{s}}(j)$.

\textit{Proof.} If $i$ and $j$ are connected directly via the left swap rule, i.e.,  $(\hat{i}_l,\hat{i}_{l+1})=(1,0)$, $(\hat{j}_l,\hat{j}_{l+1})=(0,1)$ and $\hat{j}_{l'}=\hat{i}_{l'}$, $l'\not\in\{l,l+1\}$, we have
\begin{equation*}
    w_{\mathrm{p},\mathbf{s}}(i) = \sum_{\mathclap{l'\not\in\{l,l+1\}}} \hat{i}_{l'} \beta_{\sv,l'} + \beta_{\sv,l} \le \sum_{\mathclap{l'\not\in\{l,l+1\}}} \hat{i}_{l'} \beta_{\sv,l'} + \beta_{\sv,l+1} = w_{\mathrm{p},\mathbf{s}}(j)
\end{equation*}
using Lemma 1. And if $i$ and $j$ are connected directly via the binary domination rule, 
\begin{equation*}
w_{\mathrm{p},\mathbf{s}}(i) \le w_{\mathrm{p},\mathbf{s}}(i) + \sum_{l', \hat{i}_{l'}\ne \hat{j}_{l'}} \hat{i}_{l'} \beta_{\sv,l'} = w_{\mathrm{p},\mathbf{s}}(j).
\end{equation*}
For general $i\preccurlyeq j$, the above is extended via transitivity. $\qed$

Using this $\mathbf{s}$-symmetric polarization weight, we define a rate-flexible polar code design by picking the $K$ channel indices with the largest $w_{\mathrm{p},\mathbf{s}}$. If multiple indices correspond to the $K$-th largest $w_{\mathrm{p},\mathbf{s}}$, all of them have to be included in the information set $\mathcal{I}$, potentially increasing the code dimension. This design procedure is listed in Algorithm \ref{alg:rate_comp_construction}. Note that the potential increase in code dimension does not reduce the minimum distance of the code, as indices with the same $w_{\mathrm{p},\mathbf{s}}$ have also the same Hamming weight. 

\begin{algorithm}[tbh]
	\SetAlgoLined
	\LinesNumbered
	\SetKwInOut{Input}{Input}\SetKwInOut{Output}{Output}
	\Input{Block profile $\sv=[s_0,\dots,s_{m-1}]$, $\beta$, $K$}
	\Output{Information set $\mathcal{I}$}
	$l \gets 0$\;
	\For {$k=0,\dots,m-1$}{
	    $\beta_{\mathbf{s},l:l+s_k} \gets \frac{1}{s_k} \sum_{i=l}^{l+s_k} \beta^i$\;
	    $l\gets l+s_k$\;
	}
	$\wv \gets  [ w_{\mathrm{p},\sv}(i)]_i = [ \sum_{l=0}^{n-1} \hat{i}_l \beta_{\sv,l} \; \text{ for each } i \in \ZZ_N ]$\;
	$t \gets $ $K$-th largest value of $\mathbf{w}$\;
	$\mathcal{I} \gets  \{i \in \ZZ_N | w_{\mathrm{p},\sv}(i) \ge t\}$\;
\caption{\footnotesize Polar code design based on $\sv$-symmetric $\beta$-expansion}\label{alg:rate_comp_construction}
\end{algorithm}

\textbf{Theorem 3}: All codes $\mathcal{C}$ constructed using Alg. \ref{alg:rate_comp_construction} satisfy $\operatorname{Aut}(\mathcal{C}) \supseteq \operatorname{BLTA}(\mathbf{s})$.

\textit{Proof:}
According to Theorem 1, for the proposition to hold we have to verify that the codes follow the partial order and their $P_\mathbf{s}$ stabilizes their information set.
Compliance with partial order is proven by contradiction. Assume there exists a pair $i \preccurlyeq j$ with $i \in \mathcal{I}$ and $j \not\in \mathcal{I}$. Then, from Lemma 2 we have $t \le w_{\mathrm{p},\mathbf{s}}(i) \le  w_{\mathrm{p},\mathbf{s}}(j)$, which contradicts $j \not\in \mathcal{I}$ and, thus, $\mathcal{C}$ must follow the partial order.
Due to construction, all indices within each group have the same $w_{\mathrm{p},\mathbf{s}}$, and thus, are either all included or excluded from the information set $\mathcal{I}$. As each group is stabilized by $P_\mathbf{s}$, so is their union $\mathcal{I}$. $\qed$

\section{Numerical Results}

\subsection{Code Design Parameters}
\begin{table}
    \centering
    \caption{\footnotesize Parameters of the rate-compatible polar code sequences}
    \begin{tabular}{c|ccccc}
        $N$ & 64 & 128 & 256 & 512 & 1024\\
        \hline
        $\sv$ & [1,1,1,3] & [1,1,1,4] & [1,1,1,1,4] & [1,1,1,1,1,4] & [1,1,1,1,1,1,4]\\
        $\beta$ & 1.1 & 1.1 & 1.122 & 1.134 & 1.14\\
    \end{tabular}
    \label{tab:params}
\end{table}
The proposed $\sv$ symmetric $\beta$ expansion design has two hyperparameters, namely $\sv$ and $\beta$. Their optimization is beyond the scope of this work and we assume some exemplary values for this proof of concept. For the block profile $\sv$, empirical evidence shows that the last block $s_{m-1}$ is particularly valuable for permuted decoding \cite{polar_aed, pilletPolarCodesForAED, ivanovTcom}. Intuitively, such automorphisms lead to larger permutation distances of the bit indices and, thus, to more diverse decoding results.
Therefore, we chose $s_{m-1}\le4$ and all others $s_k=1$ as a trade-off between error rate performance and a maximum dimensional granularity of at most $\binom{4}{2}=6$. It should be emphasized that the proposed code design algorithm works with any block profile $\sv$.
The value of $\beta$ is found via grid search for good \ac{BLER} performance for each blocklength $N$. The selected parameters are listed in Table \ref{tab:params}.

\subsection{Error-Rate Performance}
We evaluate the performance of rate-compatible polar codes with $64\le N\le1024$ for the \ac{AWGN} channel with \ac{BPSK} mapping. We compare the proposed $\sv$-symmetric $\beta$-expansion-based codes under \ac{AE-SC} decoding with ensemble size $M=8$ to the 5G standardized codes under \ac{CA-SCL} decoding with list size $L=8$. The 5G codes use a 6-bit \ac{CRC} code for $K<20$ and a 11-bit \ac{CRC} for $K\ge 20$ \cite{polar5G2018}. 
Note that \ac{AED} uses random automorphisms from the full affine automorphism group of each code, which may be larger than the specified minimum symmetry in the code design.
Fig. \ref{fig:k_vs_snr} shows the required \ac{SNR} to reach a \ac{BLER} of $10^{-3}$.
As a baseline, we also plot the $\mathcal{O}(n^{-2})$ approximation of the finite blocklength \ac{PPV} meta-converse bound \cite{ErsegheMetaConverse}.
As we can see, across all values of $N$ and rates considered, the proposed codes using \ac{AE-SC} perform similarly to the standardized codes using \ac{CA-SCL}, and outperform it for shorter blocklengths. However, for longer blocklengths ($N\ge512$), \ac{AED} falls behind \ac{SCL}, but still remains within 1~dB of the 5G polar code.
\begin{figure}
    \centering
    \resizebox{\linewidth}{!}{\input{tikz/K_vs_SNR_AED}}
    \caption{\footnotesize \ac{SNR} required to achieve a \ac{BLER} of $10^{-3}$ for different code dimensions $K$ and block sizes $N$.}
    \label{fig:k_vs_snr}
    \vspace{-.3cm}
\end{figure}
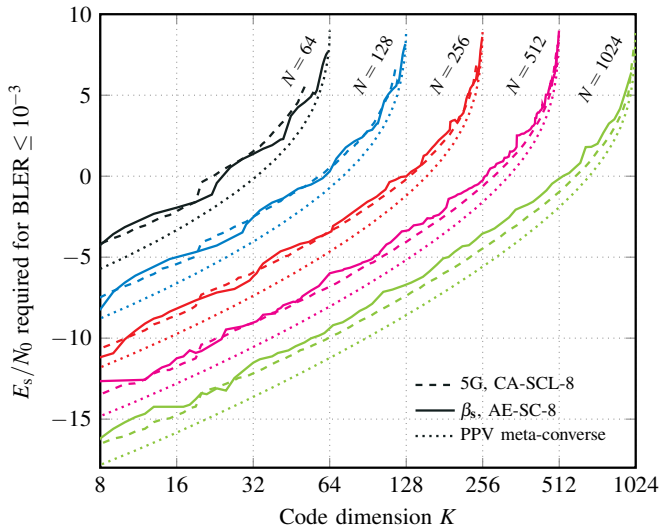
\subsection{Decoding Complexity and Latency}
As the decoders are independent, the latency of \ac{AED} is first-order independent on $M$, and its complexity scales linear in $M$. In contrast, \ac{SCL} decoding requires path management (i.e., path-metric sorting and pruning), leading to worse latency and complexity scaling. A numerical analysis of the complexity and comparison between \ac{AED} and \ac{SCL} is out of scope of this paper and we refer the interested reader to \cite{Kestel2023}.

\section{Conclusion and Outlook}\label{sec:conc}
In this work, we have investigated the necessary and sufficient conditions for generating rate-flexible polar code sequences with a given automorphism group.
Our results show that single-bit granularity is infeasible, but practical rate flexibility is achievable.
We introduced a symmetric partial order that takes both partial order and symmetry constraints into account and can be utilized for designing single polar codes or rate-compatible sequences with low complexity. 
To demonstrate our approach, we present a variant of the $\beta$ expansion to create a total order of synthetic channels that inherently ensures the desired symmetries in all designed codes.
Error-rate performance simulation results show that the designed codes using \ac{AE-SC} decoding are competitive with 5G polar codes under \ac{CA-SCL} decoding, especially in the short blocklength regime, while offering lower decoding complexity and latency.

We note that the proposed symmetric partial order can also be converted to a total order by other means of polar code design, such as density evolution \cite{moriDE} or Monte Carlo simulation restricted to the partial order graph \cite{geiselhart2022graphsearch}.

\bibliographystyle{IEEEtran}
\bibliography{references}
\end{NoHyper}
\end{document}

%% file: corporateColours.tex
\definecolor{mittelblau}{RGB}{0, 126, 198}
\definecolor{violettblau}{cmyk}{0.9, 0.6, 0, 0}
\definecolor{rot}{RGB}{238, 28 35}
\definecolor{apfelgruen}{RGB}{140, 198, 62}
\definecolor{gelb}{RGB}{255, 229, 0}
\definecolor{orange}{RGB}{244, 111, 33}
\definecolor{pink}{RGB}{237, 0, 140}
\definecolor{lila}{RGB}{128, 10, 145}
\definecolor{hellgrau}{RGB}{224, 224, 224}
\definecolor{mittelgrau}{RGB}{128, 128, 128}
\definecolor{dunkelgrau}{RGB}{80,80,80}
\definecolor{anthrazit}{RGB}{19, 31, 31}
\definecolor{darkgreen}{RGB}{34,139,34}
\definecolor{aqua}{RGB}{0, 255, 255}

%% file: nodesAndStyles.tex
\tikzset{
       vnd/.style={
        shape=circle,
        fill=black,
        draw,
        inner sep=0pt,
        minimum size=0.2cm},
        cnd/.style={
        shape=rectangle,
        fill=white,
        draw,
        minimum width=0.05mm,
        minimum height = 0.05mm}, 
         vndR/.style={
        shape=circle,
        fill=red,
        draw,
        inner sep=0pt,
        minimum size=0.2cm},
        cndR/.style={
        shape=rectangle,
        fill=white,
        draw=red,
        minimum width=0.05mm,
        minimum height = 0.05mm}
}

%% file: macros.tex
\renewcommand{\vec}[1]{\mathbf{#1}}

\newcommand{\bv}{\vec{b}}

\newcommand{\iv}{\vec{i}}
\newcommand{\jv}{\vec{j}}

\newcommand{\sv}{\vec{s}}

\newcommand{\wv}{\vec{w}}
\newcommand{\xv}{\vec{x}}
\newcommand{\yv}{\vec{y}}

\newcommand{\Am}{\vec{A}}

\newcommand{\ZZ}{\mathbb{Z}}



%% file: tikz/symmetric_partial_order.tex
\begin{tikzpicture}[xscale=0.9,yscale=.9]
\tikzstyle{code}=[circle,draw,inner sep=2pt]
\tikzstyle{code3}=[circle,draw,inner sep=6pt]

\tikzstyle{line} = [line width = 0.5mm, -latex]

\node[code] (c0) at (0,0) {};
\node[code] (c1) at (2,0) {};
\node[code] (c2) at (4,0) {};
\node[code3] (c3a) at (6,1) {};
\node[code] (c3b) at (6,-1) {};
\node[code3] (c4) at (8,0) {};
\node[code3] (c5) at (10,0) {};
\node[code3] (c6a) at (12,1) {};
\node[code3] (c6b) at (12,-1) {};
\node[code3] (c7) at (14,0) {};
\node[code3] (c8) at (16,0) {};
\node[code] (c9a) at (18,1) {};
\node[code3] (c9b) at (18,-1) {};
\node[code] (c10) at (20,0) {};
\node[code] (c11) at (22,0) {};
\node[code] (c12) at (24,0) {};

\draw[line,-latex] (c0) -- (c1);
\draw[line,-latex] (c1) -- (c2);
\draw[line,-latex] (c2) -- (c3a);
\draw[line,-latex] (c2) -- (c3b);
\draw[line,-latex] (c3a) -- (c4);
\draw[line,-latex] (c3b) -- (c4);
\draw[line,-latex] (c4) -- (c5);
\draw[line,-latex] (c5) -- (c6a);
\draw[line,-latex] (c5) -- (c6b);
\draw[line,-latex] (c6a) -- (c7);
\draw[line,-latex] (c6b) -- (c7);
\draw[line,-latex] (c7) -- (c8);
\draw[line,-latex] (c8) -- (c9a);
\draw[line,-latex] (c8) -- (c9b);
\draw[line,-latex] (c9a) -- (c10);
\draw[line,-latex] (c9b) -- (c10);
\draw[line,-latex] (c10) -- (c11);
\draw[line,-latex] (c11) -- (c12);

\node[below=2pt,align=center] at (c0.south) {00000\\\{0\}};
\node[below=2pt,align=center] at (c1.south) {10000\\\{1\}};
\node[below=2pt,align=center] at (c2.south) {01000\\\{2\}};
\node[above=2pt,align=center] at (c3a.north) {00-1-\\\{4,8,16\}};
\node[below=2pt,align=center] at (c3b.south) {11000\\\{3\}};
\node[below=2pt,align=center] at (c4.south) {10-1-\\\{5,9,17\}};
\node[below=2pt,align=center] at (c5.south) {01-1-\\\{6,10,18\}};
\node[above=2pt,align=center] at (c6a.north) {00-2-\\\{12,20,24\}};
\node[below=2pt,align=center] at (c6b.south) {11-1-\\\{7,11,19\}};
\node[below=2pt,align=center] at (c7.south) {10-2-\\\{13,21,25\}};
\node[below=2pt,align=center] at (c8.south) {01-2-\\\{14,22,26\}};
\node[above=2pt,align=center] at (c9a.north) {00111\\\{28\}};
\node[below=2pt,align=center] at (c9b.south) {11-2-\\\{15,23,27\}};
\node[below=2pt,align=center] at (c10.south) {10111\\\{29\}};
\node[below=2pt,align=center] at (c11.south) {01111\\\{30\}};
\node[below=2pt,align=center] at (c12.south) {11111\\\{31\}};

\end{tikzpicture}

%% file: tikz/table_overlay.tex
\begin{tikzpicture}

    \begin{scope}[xshift=-0.65cm]
        \draw[mittelblau, densely dashed, rounded corners] (0,-0.025) -- (3.25,-0.025) -- (3.25,0.25) --(0,.25)--cycle;
     \end{scope}
     
    \begin{scope}[yshift=-0.3cm,xshift=-0.65cm]
        \draw[apfelgruen, densely dashed, rounded corners] (0,-0.36) -- (3.25,-0.36) -- (3.25,0.235) --(0,.235)--cycle;
     \end{scope}
     
     \begin{scope}[yshift=-0.935cm,xshift=-0.65cm]
        \draw[violet, densely dashed, rounded corners] (0,-0.05) -- (3.25,-0.05) -- (3.25,0.25) --(0,.25)--cycle;
     \end{scope}
     
     \begin{scope}[yshift=-0.3cm,xshift=3.0cm]
        \draw[orange, densely dashed, rounded corners] (-0.1,-0.37) -- (1.45,-0.37) -- (1.45,0.25) --(-0.10,.25)--cycle;
     \end{scope}
     
     \begin{scope}[yshift=-0.3cm,xshift=-2.3cm]
        \draw[rot, densely dashed, rounded corners] (0,-0.37) -- (1.45,-0.37) -- (1.45,0.25) --(0,.25)--cycle;
     \end{scope}
     
\end{tikzpicture}

%% file: tikz/K_vs_SNR_AED.tex
\begin{tikzpicture}
\begin{axis}[
	width=\linewidth,
	height=.88\linewidth,
	scale=1.125,
	grid style={dotted,gray},
	xmajorgrids,
	yminorticks=true,
	ymajorgrids,
	legend columns=1,
	legend pos=south east,   
	legend style={draw=none,fill=none},
	legend cell align={left},
	xtick={0,16,...,128},
	xlabel={Code dimension $K$},
	ylabel={$E_\mathrm{s}/N_0$ required for $\text{BLER}\le 10^{-3}$},
	legend image post style={mark indices={}},
	mark size=1.5pt,
	line width = 1pt,
	xmode=log,
    xtick={8,16,32,64,128,256,512,1024},
	xticklabels={8,16,32,64,128,256,512,1024},
	xmin=8,
	xmax=1024,
	ymin=-18,
	ymax=10,
]

\addplot [color=anthrazit,dashed]
table[col sep=comma]{
4,-5.878
5,-5.323
6,-4.916
7,-4.570
8,-4.178
9,-3.832
10,-3.582
11,-3.229
12,-3.044
13,-2.764
14,-2.521
15,-2.400
16,-2.163
17,-1.935
18,-1.759
19,-1.593
20,-0.382
21,-0.196
22,-0.061
23,0.148
24,0.354
25,0.506
26,0.656
27,0.815
28,0.927
29,1.070
30,1.252
31,1.376
32,1.528
33,1.749
34,1.836
35,2.032
36,2.179
37,2.415
38,2.628
39,2.834
40,3.006
41,3.209
42,3.373
43,3.566
45,3.967
46,4.196
47,4.362
48,4.572
49,4.929
51,5.527
52,5.912
};

\addplot [color=anthrazit,solid]
table[col sep=comma]{
1 , -11.252
2 , -7.733
3 , -6.941
4 , -6.561
7 , -5.428
8 , -4.248
9 , -3.578
10 , -3.144
13 , -2.325
16 , -1.943
19 , -1.596
22 , -1.136
23 , -0.423
26 , 0.55
29 , 1.052
32 , 1.372
35 , 1.691
38 , 1.914
41 , 2.261
42 , 2.295
45 , 3.722
48 , 4.253
51 , 4.562
54 , 4.908
55 , 5.146
56 , 5.087
57 , 5.309
60 , 6.846
61 , 7.218
62 , 7.513
63 , 7.683
};
\addplot [color=anthrazit,dotted]
table[col sep=comma]{
2.0676,-9
3.4596,-8
5.1679,-7
7.3277,-6
10.044,-5
13.395,-4
17.419,-3
22.104,-2
27.365,-1
33.039,0
38.882,1
44.59,2
49.844,3
54.366,4
57.985,5
60.67,6
62.52,7
63.717,8
64,9
};

\addplot [color=mittelblau,dashed]
table[col sep=comma]{
1,-10.437
2,-9.844
3,-9.377
4,-8.989
5,-8.593
6,-8.097
7,-7.795
8,-7.480
9,-7.084
10,-6.792
11,-6.560
12,-6.248
13,-6.040
14,-5.714
15,-5.584
16,-5.425
17,-5.267
18,-5.101
19,-4.970
20,-4.102
21,-3.957
22,-3.839
23,-3.649
24,-3.541
25,-3.416
26,-3.272
27,-3.124
28,-3.052
29,-2.907
30,-2.819
31,-2.657
32,-2.492
33,-2.386
34,-2.243
35,-2.147
36,-2.033
37,-1.938
38,-1.848
39,-1.695
40,-1.602
41,-1.543
42,-1.392
43,-1.345
44,-1.234
45,-1.148
46,-1.034
47,-0.997
48,-0.836
49,-0.751
50,-0.706
51,-0.653
52,-0.553
53,-0.454
54,-0.360
55,-0.275
56,-0.183
57,-0.098
58,-0.001
59,0.084
60,0.156
61,0.230
62,0.329
63,0.433
64,0.555
65,0.594
66,0.695
67,0.739
68,0.839
69,0.943
70,1.022
71,1.134
72,1.210
73,1.318
74,1.400
75,1.524
76,1.597
77,1.648
78,1.728
79,1.799
80,1.869
81,1.981
82,2.034
83,2.157
84,2.234
85,2.361
86,2.397
87,2.498
88,2.599
89,2.661
90,2.766
91,2.854
92,2.955
93,3.040
94,3.148
95,3.233
96,3.333
97,3.471
98,3.575
99,3.715
100,3.838
101,3.982
102,4.104
103,4.212
104,4.380
105,4.508
106,4.644
107,4.777
108,4.868
109,5.025
110,5.177
111,5.334
112,5.541
113,5.803
114,6.029
115,6.304
116,6.631
};

\addplot [color=mittelblau,solid]
table[col sep=comma]{
1 , -14.263
2 , -10.718
3 , -9.996
4 , -9.472
8 , -8.226
9 , -7.148
10 , -6.57
11 , -6.142
15 , -5.117
19 , -4.685
23 , -4.281
29 , -3.476
30 , -2.981
34 , -2.116
38 , -1.63
42 , -1.369
48 , -0.779
54 , -0.519
60 , -0.128
64 , 0.294
68 , 1.088
74 , 1.771
80 , 2.15
86 , 2.405
90 , 2.767
94 , 2.865
98 , 3.239
99 , 3.257
105 , 4.668
109 , 5.067
113 , 5.353
117 , 5.582
118 , 5.874
119 , 5.796
120 , 5.918
124 , 7.509
125 , 7.761
126 , 7.976
127 , 8.162
};

\addplot [color=mittelblau,dotted]
table[col sep=comma]{
1.9525,-12
3.3757,-11
5.1471,-10
7.4456,-9
10.437,-8
14.283,-7
19.138,-6
25.134,-5
32.36,-4
40.833,-3
50.466,-2
61.041,-1
72.192,0
83.418,1
94.127,2
103.73,3
111.77,4
117.99,5
122.42,6
125.35,7
127.15,8
128,9
};

\addplot [color=rot,dashed]
table[col sep=comma]{
4,-11.851
5,-11.369
6,-11.148
7,-10.974
8,-10.640
9,-10.331
10,-10.026
11,-9.743
12,-9.513
13,-9.150
14,-8.979
15,-8.753
16,-8.611
17,-8.427
18,-8.258
19,-8.009
20,-7.318
21,-7.169
22,-7.042
23,-6.923
24,-6.802
25,-6.705
26,-6.570
27,-6.494
28,-6.273
29,-6.228
31,-6.070
32,-5.937
33,-5.793
35,-5.640
36,-5.539
38,-5.367
39,-5.278
41,-5.110
43,-4.939
44,-4.904
46,-4.699
48,-4.528
50,-4.377
52,-4.283
55,-4.062
57,-3.909
59,-3.755
62,-3.526
65,-3.383
67,-3.233
70,-3.097
73,-2.869
76,-2.707
80,-2.491
83,-2.307
86,-2.146
90,-1.937
94,-1.752
98,-1.534
102,-1.332
106,-1.161
111,-0.932
116,-0.701
120,-0.522
126,-0.281
131,-0.026
136,0.169
142,0.421
148,0.725
155,1.006
161,1.321
168,1.603
175,1.941
182,2.237
190,2.577
198,3.008
207,3.479
215,3.993
225,4.704
234,5.541
244,7.246
};

\addplot [color=rot,solid]
table[col sep=comma]{
1 , -17.215
2 , -13.736
3 , -12.94
4 , -12.501
5 , -12.149
9 , -10.934
10 , -10.07
11 , -9.544
12 , -9.103
13 , -8.794
14 , -8.64
15 , -8.364
19 , -7.669
23 , -7.184
27 , -6.964
31 , -6.436
32 , -6.191
33 , -5.992
34 , -5.655
35 , -5.397
41 , -4.641
45 , -4.422
49 , -4.178
53 , -4.058
57 , -3.791
61 , -3.621
65 , -3.408
66 , -3.073
72 , -2.695
78 , -2.292
84 , -1.929
90 , -1.584
94 , -1.43
98 , -1.26
102 , -1.03
106 , -0.83
110 , -0.364
116 , -0.227
122 , -0.05
128 , -0.014
134 , 0.282
140 , 0.458
146 , 0.664
150 , 1.257
154 , 1.342
158 , 1.554
162 , 1.758
166 , 1.931
172 , 2.214
178 , 2.426
184 , 2.626
190 , 2.86
191 , 3.356
195 , 3.431
199 , 3.474
203 , 3.318
207 , 3.648
211 , 3.747
215 , 3.912
221 , 4.786
222 , 4.878
223 , 4.978
224 , 5.037
225 , 5.033
229 , 5.426
233 , 5.597
237 , 5.812
241 , 5.983
242 , 6.17
243 , 6.333
244 , 6.195
245 , 6.464
246 , 6.376
247 , 6.479
251 , 7.781
252 , 7.959
253 , 8.215
254 , 8.399
255 , 8.6
};

\addplot [color=rot,dotted]
table[col sep=comma]{
2.9329,-22
2.2804,-21
1.5569,-20
0,-19
0,-18
0,-17
0,-16
1.8833,-15
3.3193,-14
5.1177,-13
7.4835,-12
10.621,-11
14.751,-10
20.119,-9
26.987,-8
35.622,-7
46.276,-6
59.154,-5
74.363,-4
91.865,-3
111.42,-2
132.51,-1
154.38,0
176.01,1
196.25,2
214.01,3
228.49,4
239.37,5
246.85,6
251.56,7
254.32,8
255.84,9
};

\addplot [color=magenta,dashed]
table[col sep=comma]{
4,-14.855
5,-14.554
6,-14.223
7,-13.836
8,-13.476
9,-13.141
10,-12.839
11,-12.792
12,-12.547
13,-12.177
14,-11.986
15,-11.963
16,-11.743
17,-11.550
18,-11.410
19,-11.244
20,-10.405
21,-10.282
22,-10.156
23,-9.985
24,-9.951
26,-9.688
27,-9.524
28,-9.426
30,-9.204
31,-9.133
33,-9.004
34,-8.872
36,-8.667
38,-8.450
40,-8.345
42,-8.250
44,-8.082
46,-7.932
48,-7.752
51,-7.577
53,-7.451
56,-7.228
58,-7.117
61,-6.977
64,-6.815
68,-6.613
71,-6.439
75,-6.225
78,-6.045
82,-5.850
86,-5.653
91,-5.420
95,-5.319
100,-5.092
105,-4.881
110,-4.715
116,-4.477
122,-4.209
128,-4.061
134,-3.842
141,-3.606
148,-3.400
155,-3.179
163,-2.954
171,-2.752
180,-2.492
189,-2.265
198,-2.036
208,-1.771
218,-1.514
229,-1.239
241,-0.949
253,-0.704
265,-0.391
278,-0.134
292,0.193
307,0.538
322,0.856
338,1.221
355,1.574
373,1.998
392,2.490
411,2.995
432,3.588
454,4.330
476,5.278
500,7.775
};

\addplot [color=magenta,solid]
table[col sep=comma]{
1 , -20.279
2 , -16.757
3 , -15.988
4 , -15.479
5 , -15.138
6 , -14.816
7 , -13.181
8 , -12.654
12 , -12.573
13 , -12.124
14 , -11.808
15 , -11.645
16 , -11.244
17 , -11.13
18 , -10.753
19 , -10.912
20 , -10.799
24 , -10.063
25 , -9.468
29 , -9.277
33 , -8.964
34 , -8.792
35 , -8.558
39 , -8.476
40 , -8.218
41 , -8.143
42 , -7.974
46 , -7.751
47 , -7.771
48 , -7.636
49 , -7.603
50 , -7.414
54 , -7.062
58 , -6.783
64 , -5.997
68 , -5.889
72 , -5.808
73 , -5.735
77 , -5.651
81 , -5.521
85 , -5.396
86 , -5.334
90 , -5.235
94 , -5.109
95 , -5.031
96 , -4.934
100 , -4.821
101 , -4.677
107 , -4.219
111 , -4.091
117 , -3.756
123 , -3.409
127 , -3.377
131 , -3.332
137 , -3.038
141 , -3.008
145 , -2.974
149 , -2.921
155 , -2.598
159 , -2.606
163 , -2.568
164 , -2.481
168 , -2.43
172 , -2.383
178 , -2.179
184 , -1.929
188 , -1.403
194 , -1.32
200 , -1.23
204 , -1.131
210 , -1.073
216 , -0.953
222 , -0.861
226 , -0.774
232 , -0.702
238 , -0.585
242 , -0.547
246 , -0.489
252 , -0.334
256 , -0.263
260 , -0.034
266 , 0.068
270 , 0.306
274 , 0.469
280 , 0.531
286 , 0.586
290 , 0.76
296 , 0.815
302 , 0.875
308 , 0.923
312 , 1.136
318 , 1.147
324 , 1.21
328 , 1.523
334 , 1.518
340 , 1.587
344 , 1.676
348 , 1.804
349 , 2.513
353 , 2.518
357 , 2.529
363 , 2.6
367 , 2.624
371 , 2.622
375 , 2.668
381 , 2.753
385 , 2.773
389 , 2.825
395 , 2.905
401 , 3.073
405 , 3.108
411 , 3.247
412 , 3.586
416 , 3.606
417 , 3.753
418 , 3.841
422 , 3.929
426 , 3.931
427 , 3.936
431 , 4.05
435 , 4.089
439 , 4.216
440 , 4.449
444 , 4.273
448 , 4.317
454 , 5.025
458 , 5.021
462 , 5.045
463 , 5.128
464 , 5.177
465 , 5.277
466 , 5.331
470 , 5.569
471 , 5.585
472 , 5.658
473 , 5.599
477 , 5.827
478 , 5.893
479 , 5.804
483 , 5.999
487 , 6.174
488 , 6.164
492 , 6.359
493 , 6.453
494 , 6.598
495 , 6.703
496 , 6.845
497 , 6.724
498 , 6.914
499 , 6.744
500 , 7.045
504 , 7.995
505 , 7.984
506 , 7.988
507 , 8.248
508 , 8.443
509 , 8.623
510 , 8.8
511 , 8.972
};

\addplot [color=magenta,dotted]
table[col sep=comma]{
1.8494,-18
3.2912,-17
5.1022,-16
7.5011,-15
10.714,-14
14.997,-13
20.655,-12
28.037,-11
37.548,-10
49.639,-9
64.789,-8
83.483,-7
106.16,-6
133.14,-5
164.57,-4
200.25,-3
239.61,-2
281.56,-1
324.5,0
366.38,1
404.97,2
438.23,3
464.77,4
484.15,5
497.02,6
504.79,7
509.08,8
511.32,9
};

\addplot [color=apfelgruen,dashed]
table[col sep=comma]{
4,-17.889
5,-17.367
6,-16.987
7,-16.807
8,-16.531
9,-16.208
10,-16.060
11,-15.762
12,-15.595
13,-15.184
14,-15.021
15,-14.953
16,-14.756
17,-14.593
18,-14.330
19,-14.227
20,-13.417
21,-13.220
23,-13.044
24,-12.923
25,-12.820
27,-12.591
28,-12.451
30,-12.376
32,-12.172
33,-12.077
35,-11.898
37,-11.690
40,-11.429
42,-11.317
44,-11.141
47,-10.946
49,-10.745
52,-10.635
55,-10.438
59,-10.234
62,-10.073
65,-9.875
69,-9.709
73,-9.495
77,-9.305
82,-9.029
87,-8.827
91,-8.709
97,-8.427
102,-8.236
108,-8.030
114,-7.827
121,-7.587
128,-7.390
135,-7.145
143,-6.975
151,-6.699
160,-6.471
169,-6.280
179,-6.014
189,-5.813
200,-5.590
212,-5.338
224,-5.093
237,-4.817
250,-4.567
265,-4.318
280,-4.052
296,-3.824
313,-3.511
331,-3.255
350,-3.021
370,-2.715
391,-2.408
414,-2.145
438,-1.859
463,-1.543
489,-1.232
517,-0.914
547,-0.557
579,-0.174
612,0.200
647,0.567
684,0.992
724,1.533
765,1.958
809,2.497
856,3.167
905,3.941
957,5.054
1012,8.241
};

\addplot [color=apfelgruen,solid]
table[col sep=comma]{
1,-23.377
2,-19.681
3,-19.044
4,-18.538
5,-18.178
6,-17.793
7,-17.551
8,-16.192
9,-15.600
10,-15.195
11,-14.875
12,-14.670
13,-14.238
17,-14.238
18,-14.112
19,-13.771
20,-13.399
21,-13.461
23,-13.102
25,-13.238
27,-12.462
32,-11.517
34,-11.304
38,-11.125
40,-10.918
44,-10.758
47,-10.500
51,-10.262
54,-10.013
58,-9.848
61,-9.687
66,-9.200
72,-9.021
78,-8.690
84,-8.318
89,-8.129
94,-7.882
104,-7.142
113,-7.002
119,-6.875
125,-6.758
134,-6.570
141,-6.399
149,-6.213
158,-5.880
169,-5.461
179,-5.076
188,-4.991
198,-4.685
213,-4.328
226,-4.264
240,-3.930
255,-3.551
271,-3.393
285,-3.227
305,-2.877
325,-2.545
345,-2.069
365,-1.902
384,-1.757
404,-1.577
430,-1.232
456,-0.880
480,-0.465
506,-0.212
532,0.022
560,0.190
588,0.497
620,0.822
651,1.796
685,1.812
725,2.005
764,2.679
803,3.245
844,3.511
889,4.171
935,5.251
984,6.402
};

\addplot [color=apfelgruen,dotted]
table[col sep=comma]{
1.8375,-21
3.2828,-20
5.1014,-19
7.5188,-18
10.772,-17
15.139,-16
20.954,-15
28.623,-14
38.637,-13
51.579,-12
68.131,-11
89.072,-10
115.26,-9
147.61,-8
187.02,-7
234.27,-6
289.91,-5
354.07,-4
426.27,-3
505.2,-2
588.55,-1
673.06,0
754.63,1
828.9,2
892,3
941.4,4
976.62,5
999.28,6
1012.4,7
1019.2,8
1022.6,9
};

\addlegendentry{\footnotesize 5G, CA-SCL-8};
\addlegendentry{\footnotesize $\beta_\sv$, AE-SC-8};
\addlegendentry{\footnotesize PPV meta-converse};

\coordinate (a1) at (axis cs:64,9) {};
\coordinate (a2) at (axis cs:128,9) {};
\coordinate (a3) at (axis cs:256,9) {};
\coordinate (a4) at (axis cs:512,9) {};
\coordinate (a5) at (axis cs:1024,9) {};

\end{axis}
\node[left=0.2cm of a1,rotate=60] {\footnotesize $N=64$};
\node[left=0.2cm of a2,rotate=60] {\footnotesize $N=128$};
\node[left=0.2cm of a3,rotate=60] {\footnotesize $N=256$};
\node[left=0.2cm of a4,rotate=60] {\footnotesize $N=512$};
\node[left=0.2cm of a5,rotate=60] {\footnotesize $N=1024$};

\end{tikzpicture}